\documentclass[11pt]{article}
\usepackage{geometry}
\usepackage{amsfonts,amsmath,amssymb}
\usepackage{mathrsfs,frcursive}
\usepackage{color}
\usepackage{theorem,cite}
\usepackage{hyperref}

\begin{document}
%%%%%%%%%%%%%%%%%%%%%%%%%%%%%%%%%%%%%%%%%%%%%%%%%%%%%%%%%%%%%%%
\pagestyle{empty}

\null
\vspace{20pt}

\parskip=6pt

\begin{center}
\textbf{\Large{A Century of Group Theory in Particle Physics and Beyond}}
\end{center}
\vspace{1cm}

\begin{center}
Paul SORBA
\end{center}
\begin{center}
\textsl{Laboratoire d'Annecy-le-Vieux de Physique Th{\'e}orique LAPTh, \\ 
CNRS, Université Savoie Mont Blanc, F-74940 Annecy}
\end{center}
\begin{center}
\textsl{paul.sorba@lapth.cnrs.fr}
\end{center}
\vspace{1cm}
\subsection*{Abstract}
The development of Group Theory in Mathematics as well as its impact in Physics has been spectacular during the twentieth century, and more particularly these last fifty years. If its contribution to Particle Physics deserves a special consideration, its usefulness cannot be neglected in other domains like Statistical Physics for example, and might also be of interest in Theoretical Biology. Some of these aspects will be examined.

\vspace{8cm}

to appear in the Proceedings of \emph{Nonlinearity, nonlocality and ultrametricity conference 2025} on the occasion of Branko Dragovich 80th birthday.

%%%%%%%%%%%%%%%%%%%%%%%%%%%%%%%%%%%%%%%%%%%%%%%%%%%%%%%%%%%%%%%
\newpage
\pagestyle{plain}
\parindent=0pt
\setcounter{section}{-1}

\section{Preamble}

It is for me an honour and a great chance to be a friend of Branko. May I say that we met for the first time more than fifty years ago during the Herceg-Novi Summer School of 1968? It was for both of us the first International conference we attended ! Then years passed, and it is mainly at the turn of the century that we became closer, during visits of Branko in Annecy and myself in Belgrad, in particular with our common interest to extend mathematical tools to sciences of life. I particularly appreciated his enthusiasm, his communicative energy.\\ 
As you can imagine, I was delighted to attend this International Conference on the occasion of Branko 80th Birthday, and I wish to thank the organizers Professors Zoran Rakic, Goran Djordjevic and Bozidar Jovanovic for the excellent scientific quality of the meeting and the so warm atmosphere during these days. I consider important to keep a concrete souvenir of this meeting and is happy that Proceedings will be published in honour of Branko, the man, the physicist, the humanist, the open-minded person who did, and still does, so much for science and for scientists.

\section{Introduction}

While preparing my seminar, I realized that the title I had already sent to the Organizing Committee needed some correction: indeed, it would have been more reasonable to talk about “A century” instead of “Fifty years” of Group theory in physics and beyond.  

Indeed, the predominant role of Group Theory in Particle physics increased continuously all along the last century.  Appearing today as a corner stone in this domain, it intervenes also significantly in Statistical Physics, and even to day in Genetics, without forgetting, among others, Crystallography and Chemistry with the classification of finite groups. The simple diagram:

\centerline{\textbf{nature → mathematics → symmetry → physics}}

summarizes in my opinion the famous words:

“ \textsl{The book of nature is written in the language of mathematics}” (Galileo Galilei)

“\textsl{Mathematics is just a question of groups}” (Henri Poincaré)

“\textsl{As far as I can see, all a priori statements in physics have their origin in symmetry}” (Hermann Weyl)

From a more general point of view, and following the philosopher of sciences Gaston Bachelard, in “La formation de l’esprit scientifique”, the era of a "new scientific spirit" started at the beginning of the twentieth century "at the moment when the Einsteinian relativity came to deform the primordial concepts which were considered motionless forever". It is in that period that started a spectacular development of Group Theory in Mathematics as well as in Physics, such mathematical tools being, in my opinion, perfectly adapted to the “new way of thinking”.

On the physical side, one can observe how much group theory has participated in the conceptual revolution in the ideas, in the way of thinking, through the discoveries and scientific developments during the XXth century, developments which continue at a dizzying pace to-day…

Actually, the notion of groups appeared very late. It seems that the name “group” was introduced   by Evariste Galois (1811-1832). A few years later came Sophius Lie (1842-1899) whose work on continuous groups is well-known for all of us. Let me also add the name of Elie Cartan (1869-1951) who classified the simple groups and their representations.

As far as I know, the importance of group theory in physics started at the beginning of the XXth century, the Poincaré group, itself containing the Lorentz group appearing as the symmetry group ruling the Einstein theory of Special Relativity. An essential breakthrough was also given by Emmy Noether theorem (1918) revealing a direct correspondence between continuous symmetry of the Lagrangian and conserved quantity.

A few years later, the usefulness of groups in quantum mechanics was revealed by  E.Wigner, H.Weil, J.Von Neumann . The physical community was quite hostile to that innovation. W.Pauli even coined the term “die Grouppenpest “.  Yet, it is also in these years that Pauli found out the $SO(4)$ hidden symmetry of the Hydrogen atom, his article “On the Hydrogen spectrum” being dated by January (1926), i.e. practically one century ago. One may admit that Pauli did not consider the problem in the context of group theory. In any case, any student to day knows about the Pauli matrices…

One may anyway wonder why group theory became so important in physics as well as in mathematics.  In physics, groups are useful to classify, to unify and finally to predict. As a mathematical object, one can say that a group does not come alone, but with its family of subgroups on one hand, and of representations on the other hand. Subgroups are essential to imagine the symmetry breaking - explicit, spontaneous – of a physical system and therefore new systems, while representations effectively “represent” our universe and its phenomena. 

Although finite groups, i.e. groups with a finite number of elements, constitute by themselves an important chapter in the domain of symmetry with applications in crystallography, chemistry and even in high energy physics, it is on continuous groups of transformations, namely Lie groups and their “extenr
sions”, that this report will be devoted. 

Lie groups occupy a central position between algebra, analysis and geometry. May I had that a part of their success in the physics community is also due to their property of to possess a Lie algebra? Of course topological properties of the group, if needed, have then to be considered separately. And it is indeed on their Lie algebras, that these structures have been contracted, extended (cf. supersymmetry, Kac-Moody Lie superalgebras and infinite dimensional Lie algebras, …), deformed (quantum groups, quantum affine groups,…). 

As it is well known and following Cartan classification, there are four families of simple Lie algebras, namely the orthogonal $SO(n)$, unitary $SU(n)$, symplectic $Sp(2n)$  and “exceptional” ones, this last family containing only five algebras. It is of some interest to remark that the construction of these Lie algebras can be done in terms of antihermitian matrices whose matrix elements are members of the four composition and division Hurwitz algebras, i.e. real numbers, complex numbers, quaternions and octonions. As can be expected, the non associativity of the octonion algebra imposes a particular, although unified, treatment for the five exceptional algebras $G_2$, $F_4$, $E_6$, $E_7$, $E_8$ : see the old paper of P.Ramond [1].

The plan of my talk follows more or less the development in time of the impact of group theory in particle physics and beyond. I have selected some topics which appear important to me. Of course, I cannot cover all the subjects, and the reasons are not only the lack of time, the reflection of my preferences and prejudices, I must also add my limited knowledge. 

In Section 2, Space-time Symmetry is discussed, the central point being the Poincaré group P and what I call “its family” with the “children”:  Galilei and the Carroll groups, both obtained by contraction of P in the limit of the speed of light c= infinity and c=0 respectively; the Schrödinger group; and “the parents”, i.e.  Conformal and/or (anti) de Sitter groups. 

The three following sections are devoted to what I call “Dynamics”. We start in Section 3 with a study of Hadronic spectroscopy, showing how the $SU(3)$ group of colour led to the theoretical study and experimental research of colour singlet stable multiquark states. 

It is common to say that the two pillars in development of the theory of elementary particle physics are the field theory and the group theory. A survey of Gauge Theories and its development in Supersymmetry is devoted to Section 4, while in Section 5 a general presentation of two dimensional Conformal Field Theory and Integrable Models is discussed. String and Superstring theories are naturally evoked, while too rapidly, in these two last chapters.

It would have been impossible not to mention, in these reflections on the role and the development of symmetries, the recent attempt towards “Generalized Symmetries”. This is done, although too quickly, in Section 6.  

Finally, we will leave the domain of pure physics to enter the sciences of life, showing that symmetries could have a role to play in Genetics, and proposing in Section 7, a model for the Genetic Code based on quantum groups.

\section{Space-time Symmetries: the Poincaré group and its family}

It is in 1887 that was achieved the famous Michelson-Morley experiment and in 1905 that A. Einstein proposed the theory of Special Relativity. Thus it was realized that the transformations that have the special property that the speed of light is the same in all inertial frames of reference (i.e. frames moving with constant velocity relative to each other) form the inhomogeneous Lorentz, or Poincaré group. Some decades later, Wigner result stating that any symmetry transformation can be represented on the Hilbert space of physical states by a linear and unitary operator - as long as we ignore discrete symmetry such as a reversal of the time coordinate - led to consider the classification of one particle state according to their transformations under the Poincaré group P.  This group is the semi direct product of (pseudo) rotations in the 3+1 dimensional Minkowsky space M, with metric: 
\[
g^{\mu\nu} = (1,-1,-1,-1)
\]
by translations in M. We recall that the general form of a Lie group is the semi-direct product of a semi-simple part by a radical, itself invariant subgroup of the whole: here the radical constituted by the translations is abelian. The commutation relations of the Poincaré Lie algebra are displayed in Table 1, where $J_i$, $K_i$, $i=1,2,3$ respectively denote the rotations and pseudo-rotations or boosts, and close into the simple $SO(3,1)$ algebra, and $P_i$ and $P_0$ are respectively the space and time translations. 

In parallel stands the Galilean group or invariance group of pre-Einsteinian physics, usually denoted as the non-relativistic physics group (or “Galilean relativity” group as defended by J.M. Levy-Leblond).  The Galilean group can be obtained from the Poincaré one by a (Inonu-Wigner) contraction, considering the speed of light going to infinity:  see Table 1 for a comparison between the commutation relations of the Galilean algebra with the Poincaré one. Note that the Galilean group admits a central extension. This eleventh generator, in 3+1 dimensions, is relative to the mass of the described particle – or elementary system – in the associated projective representation of this extended Galilean group. 

A second interesting contraction of the Poincaré group has been proposed by J.M. Levy-Leblond [2] and corresponds to the limit of P when the speed of light goes to zero (see again Table 1). Corresponding to an inversion of time and space with respect to the Galilean case, it has been denoted “Carroll group” as a reference to Lewis Carroll‘s book “Alice in the wonderland”. Table 1 provides with the commutation relations of the Carroll algebra and allows a comparison with respect to the Galilean ones. We will come back soon to this group since there are several attempts to develop a Carrollian physics from General relativity to statistical physics and hydrodynamics (see [3] and ref. therein).

Still in the context of contraction, it is impossible to neglect the de Sitter and Anti de Sitter groups. These groups can be defined from (anti) de Sitter spaces, themselves submanifolds of a generalized Minkowsky space of one higher dimension, with metric in the 5 dimensional case:
\[
ds^2= g^{\alpha\beta} d\xi_{\alpha} d\xi_{\beta}
\]
with $ g^{\alpha\beta} = (1,-1,-1,-1,-1)$ for de Sitter   and    $g^{\alpha\beta} = (1,-1,-1,-1,1)$ for anti de Sitter.
The metric in the (anti) de Sitter space can be written as :
\[
ds^2= \Phi^2 (x^2)  g^{\mu\nu} dx_{\mu} dx_{\nu}   \hspace{1cm}  where     \hspace{1cm} \Phi^2 (x^2)  = (1 + x^2/4\pi R^2)^{-1},   \hspace{1cm}   x^2 = g^{\mu\nu} x_ {\mu}x_ {\nu} 
\]
with $g^{\mu\nu}$  above defined. 
The $x_{\mu} $ can be seen as stereographic projection coordinates of a five dimensional pseudo-hypersphere of radius R:
\[
R^2= g^{\alpha\beta} \xi_{\alpha} \xi_{\beta}
\]
Relations between the $\xi_{\alpha}$ and $x_ {\mu}$ can be found in Ref [4]. The set of linear transformations mapping the hypersphere into itself form the $O(4,1)$ de Sitter or $O(3,2)$ anti-de Sitter group, and their infinitesimal generators $M_ {\alpha\beta}$ satisfy the commutation relations (depending the choice of the metric tensor):
\[
[M_ {\alpha\beta}, M_ {\gamma\delta}] = i (g_{\beta\delta} M_ {\alpha\gamma}+ g_{\alpha\gamma} M_ {\beta\delta} - g_{\alpha\delta} M_ {\beta\gamma} -  g_{\beta\gamma} M_ {\alpha\delta})
\]
Then, defining :
\[
\pi_{\mu} = M_{4,\mu} /R       \hspace{1,5cm}       with  \hspace{1cm} \mu = 0,1,2,3                 
\]
 from the commutation relations:
\[          
      [M_ {\mu\nu},M_ {\rho\sigma}] = i (-g_{\nu\sigma} M_ {\nu\rho} + g_{\nu\rho} M_ {\mu\sigma} +g_{\mu\sigma}  M_ {\nu\rho} - g_{\mu\rho}  M_ {\nu\sigma}) 
\]
 \[           
 [M_ {\mu\nu},\pi_{\rho}] =  i( (g_{\nu\rho}\pi_{\mu} -(g_{\mu\rho}\pi_{\nu})
 \]         
\[
      [\pi_{\mu}, \pi_{\nu} ]  = i M_ {\mu\nu}/ R^2
\]
we can see that in the limit R going to infinity (by group contraction) one recovers the Poincaré group. 
The importance of (anti) de Sitter spaces in General Relativity stand first in their property to appear among the simplest mathematical models of the universe with the observed accelerating expansion of the universe. (Anti) de Sitter space is the maximally symmetric vacuum solution of the Einstein field equations with a (negative) positive cosmological constant. 

De Sitter spaces are intensively used in quantum gravity formulated in terms of String theory or M-theory. Indeed, another important feature of anti-de Sitter space is that its boundary around any point looks like Minkowsky space. Therefore one can consider an auxiliary theory in which "space-time" is given by the boundary of anti-de Sitter space. This is the starting point of AdS/CFT correspondence [5] which states that the boundary of anti-de Sitter space can be regarded as the "space-time" for a conformal field theory (CFT). Indeed the symmetry group of a CFT is $SO(d+1,2)$ as is the symmetry group of linear transformations mapping the hypersphere of anti De Sitter into itself. This duality provides a non-perturbative formulation of string theory with certain boundary conditions. The most famous example of AdS/CFT correspondence states that type IIB string theory on the product space $AdS_{5} \times S^{5}$ (compactification of the 5 dimensions in the 10 dimensional space) is equivalent to the N=4 supersymmetric Yang-Mills theory on the 4-dim.boundary. From a “group theoretical” point of view the symmetry group underlying this theory is $SO(4,2)$. In the case of the M-theory - which unifies all consistent versions of superstring theories – two realizations have been considered : $AdS_{7} \times S^{4}$(so-called (2,0) theory in 6 dimensions) and $AdS_{4} \times S^{7}$ (N=6 super-conformal theory in 3 dimensions)[6]

Inversely, and this is another usefulness of this duality, when the fields of a quantum field theory (QFT) are strongly interacting, the ones in gravitational theory are weakly interacting and thus more mathematically tractable. Such a property has been used to study many aspects of nuclear [7] and condensed matter physics [8] by translating problems into more tractable problems in string theory.
Let us turn our attention to the asymptotic symmetries of asymptotic flat Lorentzian space-time at Null Infinity. This Null Infinity region corresponds to the terminus of all null geodesics of the type light-like (space-time interval is zero) at the boundary of asymptotically flat spacetimes (as the boundary of anti-deSitter spaces).  The symmetries at Null Infinity have been first determined by Bondi, Metzner and Sachs [9] after performing a conformal compactification on a 2-dim. sphere ( i.e. points in $E_3$ at a distance $r$ of a fixed point ) and defining a retarded time : $u$=$t-r$,  with the spherical coordinates $(r, \theta, \phi)$. Then the symmetries at Null Infinity, i.e. with r approaching infinity, are given by the BMS group, semi-direct product of the Lorentz group by an infinite dimensional Abelian one, this last one having been called “super-translation group” by the authors ( of course this has nothing to do with supersymmetry, the BMS papers dating from 1962). As could be expected, the BMS group contains the Poincaré one as subgroup, the “usual” four dimensional translation ones appearing by selecting the$I$=0 and $l$=1 spherical harmonics in the “super” generators :
\[
P_{\ell m}= Y_{\ell m} (\theta,\phi) \partial_{u}
\]

There are today several extensions, or more precisely “variations” of the BMS group [10]. In the years 2010, Barnich and Trossaert [11] proposed, instead of restricting oneself to globally well-defined transformations on the Riemann sphere --- that is global conformal transformations with associated group $SL(2,C)/Z_2$, itself isomorphic to the proper orthochronous Lorentz group --- to  focus on local properties and allowing the set of all, not necessarily invertible holomorphic mappings. In this case, Laurent series on the Riemann sphere are allowed. Their argument was that one has to focus on infinitesimal transformations without   requiring the associated finite transformations to be globally well-defined, following the attitude of Belavin, Polyakov and Zamolodchikov in their seminal paper on two dimensional conformal field theories [12]. In this way one gets two copies of the Witt algebra acting –semi direct sum- on meromorphic “supertranslations”, that is explicitly:
\[
[l_{m} , l_{n} ] = (m-n) l_{m+n}   \hspace{1,5cm}  [l’_{n} ,l’_{n} ] = (m-n) l’_{m+n}          \hspace{1,5cm}[l_{m} ,l’_{n} ] = 0
\]
\[[l_{p} , T_{m,n}] = ((p+1)/2 –m ) T_{m+p,n}   \hspace{1,5cm}              [l’_{p} , T_{m,n} ] = ((p+1)/2 –n ) T_{m,n+p}
\]
\[
[T_{m,n},T_{p,q} ] = 0
\]
Note that the supertranslations $T$ develop as meromorphic functions of the coordinates on the Riemann sphere and no more on spherical harmonics.
In the 2+1 dimensional case, the symmetry algebra of vector fields on the circle Vect ( S1), that is to only one Witt algebra,  acting on functions on the circle, while in dimensions  $(d+1)$ bigger than 4, the resulting symmetry is just the Poincaré algebra.

A relation between the BMS group and the Carroll group above mentioned or, more precisely conformal extensions of the Carroll group, have been established by Duval and coll. [13]. Their approach is based on geometric properties analogous to those of Newton-Cartan space-time manifold [14]. The authors have shown that the BMS group in 3+1 dim. (resp.2+1 dim) is isomorphic to the Conformal Carroll group in 2+1 dim (resp.1+1 dim.).
Conformal symmetry appears at the crossroads of many theories. The Poincaré group in d+1 dim. is naturally embedded in the conformal group $SO(d+1,2)$ orthogonal one, and any student in theoretical physics is aware that the largest group of space-time transformations of the free Maxwell equation, or the massless Klein-Gordon equation is bigger than the Poincaré group, being precisely the conformal group $SO(4,2)$. Just for clarity, let us remind that a basis of the $SO(4,2)$ algebra can be constructed by adding to the ten Poincaré generators (see again Table1) one dilatation $D$ and four special conformal generators $C_{i}$ and $C_0$ satisfying the commutation relations:
\[
[C_i, C_j] = [C_i, C_0] = 0  \hspace{1,5cm}   [D,C_j] = -i C_j  \hspace{1,5cm}   [D, C_0] = -i C_0
\]
\[
[D, P_j]= -i P_j    \hspace{1,5cm}           [D, P_0]= -i P_0  \hspace{1,5cm}     [D, J_i] = [D, K_i] = 0
\]
\[
[J_i , C_j] = i \epsilon_{ijk}C_k\hspace{1,5cm}[J_i , C_0] = 0   \hspace{1,5cm}      [K_i, C_j ] = -i \delta_{ij} C_0  \hspace{1,5cm}      [K_i, C_0 ] = -i C_i
\]
\[
[P_i  , C_j] = - 2i (\epsilon _{ijk} J_k + \delta{ij}D) \qquad  [P_i  , C_0] = - 2i K_i  \qquad  [P_0  , C_i] =  2i K_i \qquad  [P_0  , C_0] =  2i D 
\] 

Actually there is another interesting subgroup in the conformal one, which is the Schrödinger group, first determined by Niederer [15] as the largest kinematical symmetry group of the Schrödinger equation. Its algebra contains naturally the Galilean algebra and in addition two other generators, that we will denote $C_i$and $D$, $C_i$ for “conformal” and $D$ for “dilatation”. If $J_i$, $K_i$, $P_i$ , $P_0$  (with $i=1,..,d$)  denote as usual the rotations, boosts and space and time translations respectively, with the usual commutation relations of the Galilean algebra (see Table 1), the $C$ and $D$ generators act as follows on  the Galilean part : 
 \[ 
                  [C, J_i] = [C, K_i] =0   \hspace{1,5cm}      [C, P_i] = i K_i     \hspace{1,5cm}     [C, P_0] = - i D 
\]   
\[                 
   [D, J_i] = 0    \hspace{1,5cm}     [D, K_i] = i K_i   \hspace{1,5cm}     [D, P_i] = - i P_i   \hspace{1,5cm}   [D, P_0] = - 2i P_0     
\]
while the $C$ and $D$ commutator   reads:    
\[
[D, C] = 2 i C
\]
We realize that the three operators $P_0$, $D$, $C$ generate an $SO(2,1)$ algebra . This non compact form of $SU(2)$ joined to the dilatation action of $D$ suggests to consider this $SO(2,1)$ =$SU(1,1)$ algebra as the “conformal “ symmetry  part  of the Schrödinger algebra. This point of view can be reinforced by the property of the conformal algebra $SO (d+1,2)$ acting on the $d+1$ dim. Minkowski space to admit as a subalgebra the extended Schrödinger algebra in (d-1) +1 space-time dim. [16] itself containing the extended (with $M$ central extension: ( $[K_i, P_j] = i \delta_{ij}M$) Galilean algebra. The Schrödinger symmetry has been intensively considered in different directions, in particular in the physics of strongly anisotropic critical systems [17] and in the description of ageing phenomena [18].  An historical discussion on the conformal extension of Galilean invariance towards Schrödinger symmetry can be found in [19].  The impact of non relativistic Schrödinger symmetry has been considered in the context of AdS/CFT correspondence for the study of string theory backgrounds [20] as well as in condensed matter physics [21], see also [22].  

Moreover, comparing the commutation relations of the Carroll and extended Galilean algebra (see their commutation relations in Table 1 in which $[K_i, P_j] = 0$ in the Galilean part has been replaced by $[K_i, P_j] = i \delta_{ij} M$), it is not difficult to note that the Carroll algebra can be “formally” seen as the extended Galilean algebra in which the $P_0$ generator has been suppressed, the mass $M$ becoming the time translation $P_0$ generator in Carroll. In other words, the extended Schrödinger group contains the extended Galilean group and the Carroll one. 
  
One must add that the contraction of the Poincaré algebra towards the Galilean one can be generalized to the contraction of the conformal algebra towards the so-called Galilean conformal one, allowing in particular non relativistic versions of AdS/CFT [23].  This can be achieved by simply extending the contraction of the Poincaré generators:
\[
 K_i \to\eta K_i = K^\prime_i     \hspace{1,5cm}            and        \hspace{1,5cm}              P_i \to \eta P_i = P^\prime_i 
\]
(see Table1) to the spatial conformal generators:
\[
 C_i \to \eta C_i = C^\prime_i
\] 
In the obtained algebra the commutation relations between $K_i\prime$, $P_j\prime$ and $C_k\prime$ vanish, but the “conformal “$SO(2,1)$ algebra with generators $D$, $P_0$, $C_0$ remains. Of course this Galilean conformal algebra is different from the Schrödinger algebra, at least by the number of generators. The same kind of manipulation can be applied to the conformal algebra towards what we will call “conformal Carrollian” one by extending the contraction: 
\[
K_i\to\eta K_i = K_i^\prime \hspace{1,5cm}and\hspace{1,5cm}    P_0\to\eta P_0= P_0^\prime
\]
 to the time conformal generator:
\[
 C_0\to\eta C_0= C_0^\prime. 
\]
Then the generator $C_0^\prime$ will commute with the $K_i^\prime$, but also with $P0^\prime$, thus preventing the presence of a “conformal “ $SO(2,1)$.

Let me conclude this section by a personal remark. The particular position of the Schrödinger group, subgroup of the conformal one and itself containing the Galilean and Carroll groups, deserves all our attention, and it might also be of some usefulness in the study of the Null Infinity domain (see Ref. [24] for a developed construction of Schrödinger manifolds). More generally, non Lorentzian approaches seem to provide with interesting developments in General Relativity and beyond (see Ref [25]).\\

\begin{center}
\textbf{\Large {Poincar\'e algebra: Lorentz algebra and Translations}}

$\mathbf{SO(3,1)\rhd{T(4)}}$
\end{center}
\hspace{ 2cm}$(c\to{0})$   \hspace{ 1cm}   $ \swarrow $     \hspace{ 6cm}            $\searrow$  \hspace{ 1cm}   $(c\to{\inf})$

   \hspace{ 3,8cm}   $ \swarrow $     \hspace{ 7,2cm}            $\searrow$  
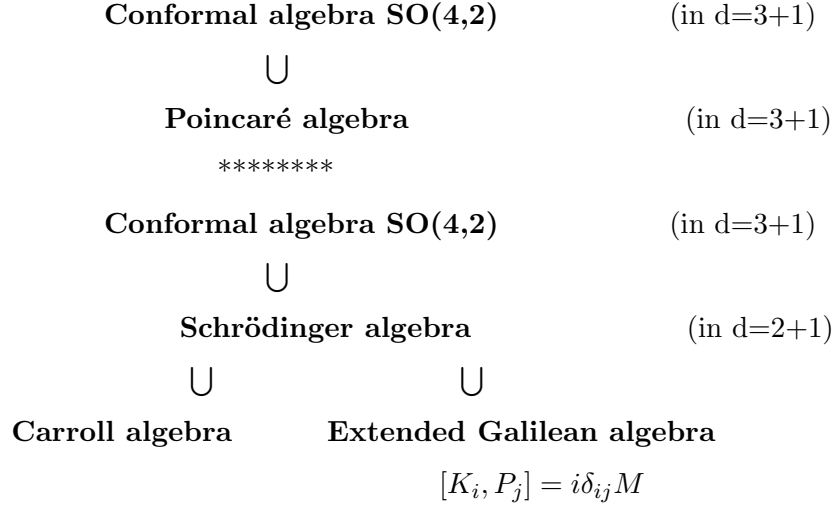
\hspace{ 1cm}   

\textbf{ \Large{Carroll algebra} }        \hspace{5,5cm}     \textbf{ \Large{Galilean algebra}}\\

$[J_{i}, J_{j}] = i \epsilon_{ijk} J_{k}$     \hspace{ 2,4cm}      $ [J_{i}, J_{j}] = i \epsilon_{ijk} J_{k} $ \hspace{ 2,6cm}  $[J_{i}, J_{j}] = i \epsilon_{ijk} J_{k}$

$[J_{i}, K_{j}] = i \epsilon_{ijk} K_{k}$     \hspace{ 2,2cm}      $ [J_{i}, K_{j}] = i \epsilon_{ijk} K_{k}$   \hspace{ 2,4cm}  $[J_{i}, K_{j}] = i \epsilon_{ijk} K_{k}$

$[K_{i}, K_{j}] = 0$     \hspace{ 3,1cm}      $ [K_{i}, K_{j}] =- i \epsilon_{ijk} J_{k}$   \hspace{2,1 cm}  $[K_{i}, K_{j}] = 0$ 

$[J_{i}, P_{j}] = i \epsilon_{ijk} P_{k}$     \hspace{ 2,4cm}      $ [J_{i}, P_{j}] = i \epsilon_{ijk} P_{k} $ \hspace{ 2,5cm}  $[J_{i}, P_{j}] = i \epsilon_{ijk} P_{k}$

$[K_{i}, P_{j}] = i \delta_{ij} P_{0}$   \hspace{ 2,5cm}      $ [K_{i}, P_{j}] = i \delta_{ij} P_{0} $ \hspace{ 2,5cm}  $[K_{i}, P_{j}] = 0$

$[J_{i}, P_{0}] = 0$     \hspace{ 3,4cm}      $ [J_{i}, P_{0}] = 0 $    \hspace{ 3,4cm}       $[J_{i}, P_{0}] = 0$

$[K_{i}, P_{0}] = 0$     \hspace{ 3,3cm}      $ [J_{i}, P_{0}] = i P_{i} $    \hspace{ 3,1cm}       $[J_{i}, P_{0}] = i P_{i}$

$[P_{i},P_{j}] =0 $          \hspace{ 3,4cm}                      $[P_{i},P_{j}] =0 $       \hspace{ 3,4cm}                          $[P_{i},P_{j}] =0 $

$[P_{i},P_{0}] =0 $          \hspace{ 3,4cm}                      $[P_{i},P_{0}] =0 $       \hspace{ 3,4cm}                          $[P_{i},P_{0}] =0 $\\

\textbf{Carroll }                            \hspace{ 3,6 cm}          \textbf{ Poincar\'e}      \hspace{ 3,8 cm}      \textbf{    Galilean}\\
 
$K_{i} \to{\eta K_{i}}$                        \hspace{ 9,9 cm}                                                     $K_{i} \to{\epsilon K_{i}}$

$P_{0} \to{\eta P_{0}}$                       \hspace{ 9,9 cm}                                                $P_{0} \to{\epsilon P_{0}}$ 

\begin{center}
\textbf{\Large{Table 1}}
\end{center}

\newpage

\hspace{5cm}\textbf{Conformal algebra SO(4,2)}    \hspace{ 2cm}      (in d=3+1)

\hspace{7cm} $\bigcup$

\hspace{5,8cm}\textbf{Poincar\'e algebra}     \hspace{ 3,4 cm}              (in d=3+1)

\hspace{6,5cm}********

\hspace{5cm}\textbf{Conformal algebra SO(4,2)}    \hspace{ 2 cm}     (in d=3+1)

 \hspace{7cm}  $\bigcup$

\hspace{6cm}\textbf{Schrödinger algebra}   \hspace{ 2,6 cm}     (in d=2+1)

 \hspace{6cm}  $\bigcup$      \hspace{ 3cm} $\bigcup$

\hspace{ 3,5 cm} \textbf{ Carroll algebra}       \hspace{ 1 cm}     \textbf{Extended Galilean algebra}

  \hspace{9,3cm}  $ [K_{i}, P_{j}] = i \delta_{ij} M $\\

\hspace{ 2 cm} Schröd. alg. =Gal. alg. + $D$, $C$ such that $\lbrace {D,C, P_{0}} \rbrace$ =$SU(1,1]$\\

\begin{center}
\textbf{\Large{Table 2}}
\end{center}

\section{Dynamics 1: Hadronic spectroscopy}

It is in the first part of the 20th century that the notion of spin and isospin appeared. The Stern-Gerlach experiment demonstrated in 1922 that the spatial orientation of angular momentum in elementary particles is quantized, then Pauli introduced the spin of the electron in 1925 and Dirac his equation for spin ½ relativistic particles in 1928. Then, in 1932 Heisenberg introduced a model for binding of the proton and the newly discovered neutron. This equal treatment of the proton and neutron being validated by experimental studies showing that these particles must bind equally, Wigner used in 1937 Heisenberg concept and introduced the term “isotopic spin”, which became isospin. As we know, the spin states as well as the isospin states of a particle are classified in representations of the $SU(2)$ group.

It is in 1961 that M.Gell-Mann and Ne’eman independently made the proposal that baryons should fall into specific representations of the $SU(3)$ “flavour” group [26]. The success of this idea, called the “eightfold way” from the title of Gell-Mann paper “The Eightfold way: A theory of strong interaction symmetry” still remains an interesting example of the way of thinking of theorists. Indeed, in this scheme, the lightest spin 1/2 and spin 3/2 baryons fall into the respectively 8 and 10 dimensional representation of the “flavour” $SU(3)$ group. But at this time, one state in the decuplet remained empty. So the same Gell-Mann proposed a year later the existence of the spin 3/2 $\Omega^{-}$. And this state was finally discovered two years later in 1964 at Brookhaven!

Soon later, in 1964, Gell-Mann and Zweig [27] independently invented quarks as constituents of hadrons. This idea might seem natural today, considering three different “flavoured” quarks of spin 1/2, belonging to the fundamental representation of $SU(3)$ and imagining the half-integer spin  baryons made of three quarks and integer spin mesons made of on quark and one antiquark. Then, due to the decomposition of $SU(3)$ representations:
\[
 3 \otimes 3\otimes 3 =1+8 +8 +10 \hspace{1,5cm} and  \hspace{1,5cm}     3\otimes \overline{3} = 1 +8
\]      
one recovers the octet and decuplet representations.

In the following year 1965, Han and Nambu [28] proposed that quarks are coloured and introduced the “$SU(3)$ colour group”, each quark belonging to the 3 dimensional representation of $SU(3)$ and hadrons being colourless, i.e. singlets of colour, a picture in perfect accordance with (Eq.1) , the singlet representation arising in the  $3 \otimes 3\otimes 3$ and $3\otimes \overline{3}$  products. One had to wait 1973 for the paper of Fritzsch, Gell-Mann and Leutwyler [29] who introduced the color octet of gluons and the Quantum chromodynamics or QCD. 

Actually, one can see that the $SU(3)$ symmetry group of fundamental importance is not the group acting over flavours, but the group acting over colour space. Indeed, it is at the basis of gauge QCD and also allows to determine the admitted hadron states, considering to other essential ingredients, namely the Pauli exclusion principle and the $SU(2)$ spin. Let us take, as an example, the case of ground state baryons and suppose we have $N$ colours $(u, d, s, c, …)$. The complete antisymmetry of the baryon wave function imposes the spin-flavour part to be completely symmetric, because of the antisymmetry in colour. Then, a simple algebraic exercice allows to determine $N(N+1)(N-1)/3$ baryons of spin 1/2 and $N(N+1)(N+2)/6$ baryons of spin 3/2: for $N=3$, one does get 8 spin 1/2 and 10 spin 3/2 particles. Note that, at least with regard to their classification, quark-antiquark mesons with a given spin and parity are simply gathered in multiplets of dimension $N^{2}$. Excited baryon states can be determined in the same way. A detailed study of this approach for the classification of hadrons and also reasonable estimates of their masses, magnetic moments, as well as the corresponding form factors relative to the semi-leptonic decays without explicit reference to the flavour unitary group, can be found in Ref [30]. 

Colour principle implies that hadrons are singlets of colour. But one can obtains states with more than three quarks satisfying this condition, and one can say that 1977 was the year where people started to bring interest in multiquark states [31].  A particularity of states with more than three quark is that there is more than one possible decomposition of such states into two colored clusters, which is not the case for an ordinary baryon (any cluster of two quarks is in the color representation $\overline{3}$ in order to combine with the third quark in the 3 dimensional color representation) or meson. This aspect was emphasized in Ref [32]] where, in order to examine the six quark content of the deuteron, it was realized that two clusters of three quarks can be either both color singlets or both color octets to constitute a singlet of color, thus exhibiting a ‘’hidden colour’’ part in such a di-baryonic system (see also [33].  Note that there is still activity on the colour structure of the deuteron in nuclear physics: see [34] for a recent study on the subject. 

This property of hidden colour has been exploited in the study of $qq\overline{q}\overline{q}$ states, also called “baryonium” [35] as well as of $qqqq\overline{q}$ ones, also called “pentaquarks” [36]. Considering the separation of two coloured clusters (for example the diquark $qq$ of colour 6 and the one of colour $\overline{6}$ in the $q\overline{q}\overline{q}$ case) by the presence, for not S-wave states, of an orbital angular momentum would prevent combinations of quarks from both sides and then would limit decays.
From an experimental point of view, the hunt to pentaquarks along the years can be compared to a real game of hide and seek. Indeed, as soon as 1978, several signals were observed in bubble chamber experiment which presented the characteristics of baryons made with four quarks and an antiquark : for a summary see [37]. But these states disappeared, and the bubble chambers were closed. Some new hope came in 2003, then in 2004 with a new narrow candidate, which itself evaporates with time.

Since my purpose is to emphasize the group theory impact in physical situations, I stay too much “light” on physical details. However, there is an important point that I cannot escape, which is the “colour magnetic interaction” [38]. Up to now, all these baryonic candidates were supposed to be made of the three light quarks u, d and s. But it was realized that the more anti-symmetry we have in flavour the stronger are the attractive forces arising from the colour magnetic interaction [39] thus reinforcing the stability of such states against decays.  Another type of argument showing that a heavy quark could facilitate the stability of a multiquark state made of two clusters separated by L is provided by considering Bohr radii [40].

And indeed it happened that the presence of charm quarks might reasonably stable produce multiquark states. The first serious indication came from the $X(3872)$  resonance reported by the Belle collaboration in 2003 , then confirmed at Babar and at the Fermilab collider (see Ref. in [41]). One possibility was to interpret this $J^{PC} = 1^{++}$ state as made of a heavy (charmed) quark-antiquark pair $Q\overline{Q}$ and a light one 
$ q\overline{q}$ [41], [42].
But we had to wait for 2015 to have the spectacular result of the LHCb Collaboration which observed two exotic structures in the $J/ \Psi$  p channel [43]. These two baryonic resonances were denoted $P_{c} ^{+}$(4380) and  $P_{c} ^{+}$(4450). Their significance is $9\sigma$ (standard deviations) for the first one and $12\sigma$ for the second one, with $J^{p}= (3/2)^ {-}$ and $(5/2)^{+}$ respectively. From expected dominant contributions, a pentaquark composition of the type ($uud\overline{c}c$) has been immediately privileged. In 2020, LHCb Collab. observed in the $J/ {\Psi}$ pair invariant mass spectrum a narrow structure at about 6,9 Gev/c which could reasonably be interpreted as a ($cc\overline{c}\overline{c}$)  tetraquark [44]: it is denoted $X$(6900). Encouraging results are obtained for these penta and tetraquarks states examined in the framework of to coloured clusters separated by an L orbital momentum barrier [40]; on a different point of view see also [45]. 

I didn’t resist in talking too much on multiquarks: sorry! But lot of efforts are still in progress as well as by theorists as by experimentalists to write this new page of hadron spectroscopy in which, as we could see, technics of group theory are essential due in particular to the fundamental role of the $SU(3)$ colour group.

\section{Dynamics 2: Gauge Theories and Supersymmetry}

The advent of local gauge theories in particle physics appears to me as a perfect example of the relationship between science and philosophy, since it leads to the idea of “unification” of fundamental forces. And this progress is deeply connected with the notion of symmetry and group theory.
It is in 1954 that C.N. Yang and R. Mills made this brilliant suggestion: “We define isotopic gauge as an arbitrary way of choosing the orientation of the isotopic spin axes at all space-time points, in analogy with the electromagnetic gauge” [46]. 

Considering the electro-dynamical theory with minimal coupling:
\[
 \Phi^{\star}(\partial_{\mu} – I. e A_{\mu})\Phi
\]
invariant under the Abelian $U(1)$ symmetry  with a  “x-dependence” of the group element parameter $\theta(x)$ and the field $\Phi$ of charge q transforming as:
\[
\Phi(x) \to U(\theta(x)) \Phi(x) = \exp(-i q\theta(x)) \Phi(x)
\]
the introduction of the covariant derivative :
\[
D_{\mu}=(\partial_{\mu} – i. e A_{\mu})
\]
with I representing the identity generator of the $U(1)$ Lie algebra, appears naturally. Indeed, under the transformation  
\[
\Phi\to\Phi^ \prime = U(\theta(x))\Phi
\]
 one gets:
\[
(\partial_{\mu} –ieq I.A_{\mu} )\Phi(x) \to (\partial_{\mu} –ieq I.A_{\mu}^\prime) \Phi^\prime(x)
\]
that is $D_{\mu}\Phi(x)$ transforming like $\Phi(x)$: 
\[
D_{\mu}\Phi(x) \to  U(\theta(x))D_{\mu}\Phi(x) 
\]
 with the connection -or gauge field- $A_{\mu}$ transforming as:  
\[
A_{\mu}\to A_{\mu} - (1/e) \partial_{\mu} \theta(x)
\]
Now generalizing the local gauge invariance to a non –Abelian Lie group $G$ with generators $X_i$with $i=1,...,n$ :
\[
\Phi(x) \to  U(\theta(x))\Phi(x)  =  \exp(-i X_i \theta^i(x))\Phi(x)
\]
the associated covariant derivative becomes :
\[
D_{\mu}= ( \partial_{\mu}  –ig X_i.A^i _ {\mu})
\]
the coupling g, analogous to e being arbitrary, and the n  gauge fields transforming as: 
\[
X_i.A^i _{\mu} \to  U(\theta) X_i.A^i _{\mu} U(\theta)^{-1} - i/g (\partial_{\mu} U(\theta)) U(\theta)^{-1}
\]
 One could say that the local Abelian gauge invariance was “hidden” in the electro -dynamical theory due to the hidden $U(1)$ generator, and it was owing to the wonderful idea of Yang and Mills, who considered  the $SU(2)$ isotopic spin, that the generalization to the non-Abelian case led to the unification of the electromagnetic, weak and strong interactions with the $SU(2)\times U(1) \times SU(3)$ gauge group in the seventies.  Started in that period, the experimental confirmation of the existence of quarks, of weak neutral $Z$ and charged $W^{+,-}$ currents, and finally of the Higgs particle in 2012 allowed to validate what is known to-day as the Standard Model of particle physics describing three of the four fundamental interactions and classifying all known elementary particles [47]. It is worth to add that the Higgs model [48] is also technically based on symmetry … and on –spontaneous- breaking of symmetry. I will not say more on this point, but just recall the words of Philip Anderson: “It is only slightly overstating the case to say that physics is the study of symmetry”.

In the second part of the seventies, encouraged by the success of the $SU(3 )\times SU(2) \times U(1)$ gauge group, theorists tried to englobe it in a simple group, and $SU(5)$ and $SO(10)$ appeared as the favourite ones for such a “Grand Unification Theory”. But the lack of experimental evidence of the proton decay, which was required in such models, stopped the hopes in this direction. 

More spectacular was in the same period the advent of Supersymmetry. One could say that two different questions converged at this period. The first one consisted in finding a consistent way to relate bosons and fermions while the second one dealed with the possibility to combine nontrivially internal and space-time symmetries.

First attempts came in 1971 to determine supergauge transformations in dual models, specially in their formulation as two-dimensional field theories [49]. Then, three years later supergauge transformations were defined in four dimensional space-time dimensions [50]. Concerning the second problem, the so-called “no-go theorem “of Coleman and Mandula (1967) reduced the possibilities of space time and internal symmetries of the $S$ matrix in quantum field theory to be represented by the direct product of the Poincaré group by a internal symmetry one [51]. It was remarked in 1971 that this restriction can be suppressed by extending the direct sum of:
\[
\mathcal{B= P+A }
\]
where $\mathcal{P}$ is the Poincaré algebra and $\mathcal{A}$  an internal symmetry one, into a “superalgebra”[52], or more precisely a $Z_2$ graded algebra with “bosonic “ generators in $\mathcal{B}$ and fermionic ones in $\mathcal{F}$ satisfying globally:
\[
\mathcal {[B,B] = B    \hspace{1,5cm}               [B,F] = F    \hspace{1,5cm}          \{ F, F \} = B}
\]
the anti-commuting fermionic generators $\mathcal{F}$ undergoing the action of $\mathcal{P}$ as well as of $\mathcal{A}$, in other words being the “link” between $\mathcal{P}$ and $\mathcal{A}$ as we will see in the next lines. 

Indeed, such an approach has been generalized in 1975 by Haag, Lopuszanski and Sohnius [53] who provided a general study of possible symmetries of the S matrix in four dimensions. The super-symmetry algebra will include as “bosonic” (or “commuting”) generators the $M_{\mu\nu}$ and $P_{\mu}$ generators of the Poincaré algebra $\mathcal{P= L + T}$ and the generators $X_A$ of the internal symmetry algebra $\mathcal{A}$, and in addition “fermionic” (or “anti-commuting”) spin 1/2 Hermitian and anti-Hermitian generators:
\[
\mathcal{Q} = \{{Q_{\alpha}^A, \bar{Q}_{\dot{\alpha}}^A}\} \hspace{1,5cm} with\hspace{1,5cm} A= 1,…, N.
\]
More precisely the Weyl spinors $Q_{\alpha}$ and their Hermitian conjugate  $\bar{Q}_{\dot{\alpha}}$ are elements of the representation (1/2, 0) and (0, 1/2) of the Lorentz part $ \mathcal{L}$, providing by anti-commutation the translation generators $P_{\mu}$ , that is a Majorana spinor in the  (1/2,1/2) Lorentz representation, while the $Q^A$ and $\bar{Q}^A$  undergo the action  of the $\mathcal{A}$ algebra. In other words, the bosonic part $\mathcal{B}$ is the direct sum of the space-time $\mathcal{P}$ and internal $\mathcal{A}$ algebra:  
\[
\mathcal{B = P + A}  \qquad \text{that is} \qquad   \mathcal{[P, A]} = 0
\]
and:                                    
\[
 \mathcal{\{ Q, Q\} }=  \mathcal{T} + \text{possible central extensions} \;\; Z
\]

The direct product of the space-time and internal algebras is still respected, following the Coleman-Mandula theorem, but now $\mathcal{P}$  and $\mathcal{A}$  are “related” by fermionic generators.

A more precise study of these supersymmetry algebras will lead to the impossibility to obtain an algebra A equal to or containing the $G= SU(3 )\times SU(2) \times U(1)$ group of interaction for massive elementary particles.  Indeed, to get renormalizable theories N must be less or equal to 4 but then $\mathcal{A}$ is not big enough not contain $G$. Among the difficulties to deal with extended supersymmetries, are the constraints, already in N=2, which arise for extending the off-shell hypermultiplet of matter with spin less of equal to 1/2 [54] [55]. We can mention at this point the harmonic superspace approach developed to remedy this situation [56]. Let us add that N=8 is the largest value for an admissible supergravity theory. 

Elementary particle physicists rapidly preferred to adopt “a variant” of this program by limiting themselves to the case N=1, but with as internal gauge algebra the recognized  $ SU(3 )+ SU(2) + U(1)$ (the “+” being for direct sum) one  not acting on the fermionic generators $Q_{\alpha}$ and $\bar{Q}_{\dot{\alpha}}$ [57]. Note also some attempts to consider in the same spirit the case N=2 [58]. In this approach, usual fields become the components of superfields [59], the $Q_{\alpha}$ and $\bar{Q}_{\dot{\alpha}}$  changing a boson into a fermion and vice-versa.  In such a framework, to each constituent of matter is associated a mirror partner, to each quark of spin 1/2 is associated a “squark” of spin 0, and so on for each particle and interaction gauge boson.  Up to now, there is no experimental evidence confirming this model usually called “Minimal Supersymmetric Standard Model”.

It is of course a natural question to ask whether nature allows to treat “on the same footing” bosons and fermions, although they are submitted to different laws (Pauli principle only for fermions) and one could say that the above approach proposes a “super-unification” of electroweak and strong interactions. But increasing the symmetry leads naturally to increase the number of particles (the bigger the group, the bigger its representations) and it is what happens in the above case, in which to each elementary fermion (boson) must be associated an elementary boson (fermion).   

We will not go farther in the study of possible supersymmetries. Our purpose is to show the emergence of Lie superalgebras, and more precisely now of simple Lie superalgebras. By chance, it is at the same period (1975-77) that V. Kac generalized the Cartan construction of simple algebras to “$Z_{2}$ graded” simple ones [60]. As a rapid survey, one finds: the unitary $SU(m|n)$ superalgebras with “bosonic” part $SU(m)+ SU(n)+U(1)$ and “anticommuting” part the representations $(m, \bar{n}) + (\bar{m},n)$
 of the $SU(m)$ and $SU(n)$ respectively, the orthosymplectic superalgebras $Osp(M|2n)$, M= 1,2, 2m, with $O(m) +Sp(2n)$ “bosonic” part and (m,2n) representations of the orthogonal and symplectic algebras , to which must be added three “exceptional” superalgebras namely $G_{3}$, $ F_{4}$, and $D(2,1; \alpha)$. A more detailed description would of course be necessary [4], in particular the unitary algebras may appear in their non compact form and, in the case of $Osp(m|2n)$ ones the orthogonal and symplectic part cannot appear simultaneously compact. Let us simply remark that the anticommuting parts stand as the fundamental, that is the simplest non trivial, representations of the bosonic algebras. Thus simple superalgebras appear as a direct generalisation of the “physical” superalgebras above introduced, their particularity standing mainly in the anticommuting part  $\mathcal{\{ Q, Q\} }$ which now reproduces the bosonic part.

A simple way, at least for me, to realize that simple Lie superalgebras naturally show up is to take as an example the case of massless particles. Then the space-time symmetry in 3+1 dimensions becomes the conformal group $SO(4,2)$. We can recognize in its Lie algebra the Poincaré one, semi-direct sum of Lorentz algebra $L$ by the translations $T$ , but also the semi-direct sum of $L$  on the four special conformal generators $K$.  Now, on the same way a first set $\mathcal{Q} = \{{Q_{\alpha}, \bar{Q}_{\dot{\alpha}}}\}   $ can be added for the part $\mathcal{ P= L + T}$ , a second set $\mathcal{Q^\prime} = \{{Q_{\alpha}^\prime, \bar{Q}_{\dot{\alpha}}^\prime}\}  $    can be added for the part $\mathcal{P= L + K}$. Then one can see that, if the anticommutation of the $\mathcal{Q}$ (resp.  $\mathcal{Q^\prime}$) generators between themselves provide the $\mathcal{T}$ (resp. $\mathcal{K}$)ones, the anticommutation of the $\mathcal{Q}$ with  $\mathcal{Q^\prime}$ close into the bosonic part of the simple superalgebra $SU(2,2|1)$!

As mentioned in the introduction, we will not give the place they deserve to superstring and supergravity theories in this communication, leaving this program for a future project [61]. Let us just mention that in 10-dimensional Type IIB supergravity with space-time manifold $AdS_{5} \times S^{5}$ (see Section 1) the associated superalgebra is $PSU(2,2|4)$ --- that is the quotient of $SU(2,2|4)$ by the $U(1)$ commuting with $SU(2,2)+SU(4))$. In the same way, in the 11 dimensional M-theory with space-time manifold $AdS_{7} \times S^{4}$, the corresponding superalgebra is $Osp(8^\star/4)$ with bosonic part $SO(6,2)+ Sp(4)$ --- we recall $Sp(4) = SO(5)$ --- while with space-time manifold $AdS_{4} \times S^{7}$, the corresponding simple superalgebra is $Osp(8/ 4,R)$ with bosonic part $ SO(8) + SO(2,3)$ [62] Let us add that these superalgebras correspond to the maximal number of supersymmetries for these theories (see ref.[63] for a general view on superstrings and supergravity and also ref.[64] on the supersymmetric gauge theories and the AdS/CFT correspondence). As a last word on this huge subject, let us mention the present attempts to use string theory for the understanding of the dark energy [65].

\section{Dynamics 3: From 2-dim. Conformal Field Theory to Integrable Models}

Among the interesting theoretical developments in the eighties, the year 1984 was particularly important with the production of two papers. The first one, already mentioned [12] concerns the modern study of conformal invariance in two dimensions, while the second one [66] brought major simplifications in open superstring theories. Although in distinct areas, the new symmetry concepts proposed in the first paper are widely used in the second. In Ref [66] was shown that the open superstring is anomaly free if and only if the gauge group is $Spin(32)/Z_{2}$. In addition the authors found that the ten dim. supersymmetric Yang-Mills field theory is anomaly free for the gauge group $Spin(32)/Z_{2}$, and also for the phenomenologically more interesting group $E_{8}$ x $E_{8}$ which is however excluded in the open string theory. This result, called the Green-Schwarz anomaly cancellation mechanism is considered as the starting point of the “first superstring revolution in superstring theory”, the “second superstring revolution” starting at the end of the twentieth with the M- theory, that is a theory that unifies all consistent versions of superstring theories and conjectured by Witten [67]. In Ref [12] the authors have in particular shown that the two dimensional conformal field theories (CFT) can be treated from a group theoretical point of view. 

Indeed, in their seminal work, the BPZ team, i.e. Belavin, Polyakov and Zamolodchikov, combined the representation theory of the Virasoro algebra (see further) with the idea of an algebra of local operators $\phi$’s and shown how to construct completely solvable conformal theories. The dynamical principle here is the associativity of the operator algebra (bootstrap hypothesis). The key ingredient is the assumption that the product of local quantum operators can always be expanded as a linear combination of well-defined local operators:
\[
\Phi_{i}(x) \Phi_{j}(y) = \sum_{k}C_{ij}^{k}(x-y) \Phi_{k}(y)
\]
 where $C_{ij}^{k}$ is a c-number. Introducing this essential property of operator product expansion will be useful in the example provided in the short section on Generalized Symmetries which follows.

From a group theoretical point of view, this approach started the uses and development of infinite dimensional groups in physics. Of course, some infinite symmetries already appeared in physics with the BMS group in General Relativity: see our previous section. But it is with ref[12] that their development really started. Indeed, it happens that, in the two dimensional case (i.e. 1 space and 1 time), any analytic mapping of the complex plane onto itself is conformal. It follows that the algebra of the “local” conformal group is obtained by extending the two $SO(2,1)$ algebras constituting the “global” conformal algebra $SO(2,2) = SO(2,1)+SO(2,1)$ , into two copies of the de Witt algebra:
\[
[l_{m} , l_{n} ] = (m-n) l_{m+n}  \hspace{1,5cm}   [l’_{n} ,l’_{n} ] = (m-n) l’_{m+n}      \hspace{1,5cm}    [l_{m} ,l’_{n} ] = 0
\]

${l_{1},l_{0},l_{-1}}$ as well as ${l_{1}^{\prime},l_{0}^{\prime},l_{-1}^{\prime}}$ constituting the two “global conformal” $SO(2,1)$ algebras.

Due to conformal invariance, the stress energy tensor $T(z)$ in a two dim.CFT can be developed with the mode operators $L_{m}$ and $L_{n}$ satisfying:
\[
[L_m , L_n ] = (m-n) L_{m+n} + (c/12) m(m^{2}-1)\delta_{m+n,0}    
\]
and
\[
[L’_m , L’_n ] = (m-n) L’_{m+n} + (c/12) m(m^{2}-1)\delta_{m+n,0}  \quad \text{with} \quad [L_m ,L’_n ] = 0
\]
$c$ being the central charge of the theory. This is the famous Virasoro algebra.

This “bootstrap approach” above mentioned is powerful in the case of theories with a finite number of representations of the Virasoro algebra, in other words a finite number of conformal families. Such so-called “Minimal Models” were identified with various two-dimensional statistical models, e.g. Ising, Potts and so on, at their critical point: for a detailed presentation see [68]. From an algebraic point of view, it is worth to mention that a classification of minimal conformal invariant theories can be obtained in terms of the simply laced algebras or A-D-E classification [69].

Is it necessary to recall that a string describes in time a two dimensional world sheet in a d dim. Minkowski space, instead of a world line for an ordinary particle in a four dim. Minkowski space? It then happens that the string action in the conformal gauge is invariant under two dim. conformal transformations with the associated Virasoro algebra. Thus string scattering amplitudes are expressed in terms of correlation functions of a CFT defined on a plane (tree amplitudes), on the torus (one-loop amplitudes) or some higher genus Riemann surface. Physical string states correspond to primary fields of the conformal field theory,i.e. fields $\Phi^{\prime}s$ transforming as follows under a conformal transformation:  
\[
\Phi(w,\bar{w}) = (dw/dz)^{-h} (d\bar{w}/d\bar{z})^{-\bar{h}} \Phi(z,\bar{z})
\]
$h,\bar{h}$ being the homomorphic and anti-homomorphic conformal dimensions and  $w$ and $z$ complex numbers. 

Conformal invariance lead rather naturally to the presence of infinite dimensional Lie algebras constructed from simple Lie algebras and called affine Kac-Moody (KM) algebras [K2]: starting form a simple complex Lie algebra $G$ and considering the ring of Laurent polynomials $C(z,z^{-1})$ the complex variable $z$  with   l$z$l = 1 (i.e. on the circle)  one can extend $G$ generated by the $T_{a}$ ($a$= 1,…,n)  with :  
\[
[T_{a}, T_{b}] = i f_{ab}^{c} T_{c}
\]
to the algebra with generators:  $z^{m}. T_{a}= T_{a}^{m}$   satisfying the commutation relations: 
\[
[T_{a}^{m}, T_{b}^{n}] = i f_{ab}^{c} T_{c}^{m+n}  - (k/2) m \delta_{m+n}< T_{a} ,T_{b}> e
\]
where $< , >$ is the Killing form on $G$
 and $e$ the central extension: we will call $G(S^{1})$
 the above Kac-Moody algebras and more generally Affine Algebras. 

Kac-Moody algebras appear in two-dimensional current algebras when considering a massless spinor (or scalar field), such a model leading to an exactly solvable CFT (for a review on the subject see [70]. They are the two-dimensional expression of gauge symmetry in string theory: for ex., in the heterotic string model, string states are classified in a representation of the KM algebra of the gauge group $E_{8}\times  E_{8}$ or  $Spin (32)/ Z_{2} $ [71]. The best example of the deep connection between a KM algebra and the Virasoro algebra is doubt given by the Wess-Zumino-Witten (WZW) models [72] where both algebras are symmetries of such field theories. 

In the same way can be defined Super Kac-Moody algebras starting from a simple super algebra (see for ex. [73]). Extension to Kac-Moody algebras defined on $M \times G_{N}$, $M$ being a smooth closed compact manifold and $G_{N}$ the Grassmann algebra with N generators have also been considered in some detail [74]and more recently with $M$ a non compact manifold, coset spaces and soft deformations [75].

At this point, it is natural to mention the algebraic Sugawara construction allowing a realization of the Virasoro algebra from any KM one. The procedure consists in embedding $G(S^{1})$ in its enveloping algebra and defining a Fock space based on a vacuum $|0\rangle$ such that: $T_{a}^{m} |0\rangle = 0$ if $m> 0$. Then one can define a normal ordering for the product   $:T_{a}^{m}. T_{a}^{n}:$ leading to the Virasoro generators $L_{m}$ in terms of $T_{a}^{m}$  :
\[
L_{m}= (1/k +\lambda)  \sum_{a=1}^{dimG} \sum_{n}:T_{a}^{n} \, T_{a}^{m-n}:
\]
$ \lambda$ being the Casimir of the adjoint representation of $G$, and such that:
\[  [L_m , L_n ] = (m-n) L_{m+n} + (c/12) m(m^{2}-1)\delta_{m+n,0}  \hspace{1.5cm}   [L_{m} , T_{a}^{n} ]  = -n T_{a}^{m+n}
\]
with : $c= k.dim(G)/ (k + \lambda)$

The supersymmetric extension of the Virasoro algebra will naturally be obtained by adding to the conformal spin 2 stress energy tensor $T(z)$ a conformal spin3/2 fermionic field $G(z)$. Then developing in Laurent modes such that: $G(z) = z^{-3/2-r}G_{r} $  with $r$ integer (Ramond sector) , $r$ half integer (Neveu-Schwarz sector), one gets the commutation relations of the N=1 Superconformal Algebra.
\[
        [L_{m} , G_{r}] = ((1/2)m –r) G_{m+r}    
\]
\[
           [ G_{r},  G_{s} ] = 2 L_{r+s}+ c/3 (r^{2}-1) \delta_{r+s,0}
\]
Another important generalisation of the Virasoro algebra is given by the so-called $W$-algebras, and super $W$-algebras. Let us restrict to the classical case. Then a classical finitely generated Wn algebra will be defined as a Lie algebra with a Poisson braket $\{. , .\}_{PB}$ and a set of generators involving a stress energy tensor $T$ as well as a finite number of primary fields $W_{h_{i}}$(i=1,…,n-1) under $T$ satisfying: 
\[
\lbrace{T(z) , T(w)}\rbrace _{PB}= -2 T(w)\delta^\prime (z-w) + \partial{T}(w) \delta(z-w) +(c/2) \delta^{\prime\prime\prime}(z-w)
\]
\[
\lbrace{T(z) , W_{h_{i}} (w) }\rbrace _{PB} = -h_{i} W_{h_{i}} (w)\delta^\prime (z-w) + \partial{W_{h_{i}}} \delta(z-w)
\]
\[
\lbrace{W_{h_{i}}(z) ,W_{h_{j}} (w) }\rbrace_{PB}  = \sum_{a} P_{i,j;a} (w) \delta^{(a)}(z-w)
\]
where $P_{i,j;a}$  are polynomials in the primary fields $W_{h_{j}}$, $T$ and their derivatives. These extra symmetries, bigger than the conformal ones, could help to characterize degeneracies, and to classify in a simpler way the physical states. $W$ algebras appear in the quantum Hall effect, black holes, in lattice models of statistical mechanics at criticality, and in Toda theories [76] as symmetry algebras [77]. 

Mentioning Toda theories project us in the domain of integrable models, the mathematical development of which arises almost at the same period and in direct connection with 2 dim. CFT and SuperString theories. Among them Toda theories stand in a particular position. Indeed, the constant of motion of a Toda theory form a $W_{n}$ algebra, and such a Toda theory can be seen as a gauged WZW model on which constraints have been imposed. In this framework, it is possible to explicitly construct, from a given simple algebra $G$, a different$W_{n}$ algebra for each (non conjugate) $Sl(2)$ embedding in $G$. Super $W_{n}$ algebras can also be constructed, then starting form a simple super algebra $SG$. Their classification and construction will be associated with the different $Osp(1|2)$ –natural super symmetric extension of $SU(2)$ – in the $SG$ under consideration. A summary on General properties of classical $W$ algebras can be found in [78].

I wish to add another remark concerning the Toda models. A Toda field theory is related to a simple algebra $G$ or affine simple algebra $G(S^{1})$: we will then call it an affine Toda field theory, the corresponding Lagrangian being : 
\[
L = 1/2 (\partial_{\mu} \Phi , \partial_{\mu} \Phi )   - ( m^{2}/ \beta^{2} ) \sum_{i=0}^{r} n_{i} . \exp(\beta.\alpha_{i} ,\Phi)
\]
Here $\Phi = (\Phi_{1},...,\Phi_{n})$ is a n-component field transforming as scalars under the Lorentz group while $m$ and $\beta$ define a classical mass scale and coupling constant, respectively. The $\alpha_{i}$ are the simple roots of the Lie algebra $G$ or $G(S^{1})$ of rank $r$ under consideration (note that $ r=1,…,n$ for $ G$ and $ r=0,1,…,n$) for $G(S^{1})$, and the inner product $(.,.)$ being the Killing form relative to $G$ or $G(S^{1})$. Finally, the ni are integers known as Kac labels or Dynkin labels. Let us add that the best known examples of affine Toda field theories are the Sine-Gordon and Sinh-Gordon models for imaginary and real values of the coupling constant, the algebra of consideration being the affine $SU(2)$ one. The generalisation from the initial Toda model to a general one [79] associated with a Lie algebra or affine Lie algebra arises by using the Killing form of an algebra of rank $n$ or affine algebra of rank $n+1$ and considering not only one but $n$ scalar fields $\Phi_{i}, i= 1,…,n$. Such a generalisation could be compared, all things considered, to the generalisation in gauge theories (see Section: Dynamics 2) when going from the Abelian $U (1)$ case to a simple Lie algebra or affine Lie algebra of rank bigger than one.

As already mentioned, there exist strong connections between integrable systems and supersymmetric Yang-Mills theories: see for ex. [80] for a review on the correspondence between the Seiberg-Witten theory [81] and the elliptic Calogero-Moser integrable systems [82]. Let us add en passant that Toda systems may also be obtained as a limit of elliptic Calogero-Moser systems…

It may be a good place to mention now the so-called Quantum Groups, which appeared in the works of V. Drinfeld on one hand and of M. Jimbo on the other hand in the middle of the eighties. I will limit myself to a very simplified and incomplete presentation of such important mathematical objects (see ref. [83]). Let us consider the universal enveloping algebra $U(G)$ of a Lie algebra $G$, i.e. the space of polynomials and formal power series in the elements of $G$ on which we apply the commutation relations appropriate for this Lie algebra. Then the quantum group $U_{q}(G)$ will be a deformation relative to the parameter $q$ of $U(G)$. As an example, considering $U_{q}(Sl(2))$, the corresponding C.R. will be:
\[
          [J_{+}, J_{-}] = \frac{(q^{2 J_{3}} - q^{-2 J_{3}} )} {(q - q^{-1}) } \hspace{1,5cm}     [J_{3}, J_{+}] =  +J_{+}     \hspace{1,5cm}              [J_{3}, J_{-}] =  -J_{-}  
\]
In the limit $q=1$, one gets back the usual $Sl(2)$ algebra, with : $ [J_{+}, J_{-}]= 2 J_{3}$. Another important limit is $q=0$, providing “Crystal basis representations” of $U_{q=0} (G)$. It appears that states in the product of representations of $U_{q=0} (G)$ do not show up as linear combinations of states, each constituted by product of states of the representations one start from, as is the case for ordinary simple algebras, but by only one ordered state of such basic states. This property is at the basis of model of the genetic code rapidly presented in our last section.

The importance of Quantum groups stands in particular in Statistical Models. For example, the symmetry of the quantum Heisenberg model denoted “XXZ”, actually a deformation of the “XXX” model, is the affine quantum group $U_q(\widehat{Sl}(2))$, a useful property to prove its integrability. Let us just add that the symmetry of the XXX model is the Yangian of $Sl(2)$ or $Y(Sl(2))$, Yangian being another type of quantum group (see again [83]). As far as I know, the symmetry of the XXZ model cannot be seen as a breaking of the symmetry of XXX model. 
 
 Extensions of the notion of Quantum groups have been considered in different directions, such as super quantum groups $U_{q}(SG)$ or  the  already mentioned affine quantum groups $U_{q}(\widehat{G})$,  Algebras with two deformation parameters  as the elliptic algebras $A_{p,q}(\widehat{Sl}(N))$ and $B_{q,l}(\widehat{G})$ can be defined from the important Yang-Baxter equation, a consistency equation to prove the integrability of a system  [84]. Note also that deformed Virasoro and $W$ algebras can be obtained from quantum elliptic algebras [85]…

As a conclusion of this rapid survey on the impact of group theory in the development of Integrable Models, let me pick up the words of the authors of ref. [76]: “\textsl{All the experience accumulated in the work with integrable systems makes it clear that the integrability properties are defined by the structure of their internal symmetry algebra}”.

\section{Generalized Symmetries}

There is some excitement these recent years about what are called “Generalized Symmetries” [ 86] I don’t think able to present a general landscape of this domain in which geometry, particularly differential geometry plays an important role (for reviews see:  [ 87]).What I can say is that the notion of group is more or less disgarded, and I will limit myself to present a special case of such of “new type” of symmetries, in which the invertibility of an element is lost (see for ex. [88].  It is the case of the so-called “fusion rules” in 2 dim.CFT (see Ref. [89]).

Let us recall that a primary field $\Phi$ of a minimal theory, that is a theory with a finite number of primary fields, corresponds to the highest weight of a Virasoro representation. Deriving the fusion of two such fields $\Phi_{1}$ and $\Phi_{2}$ stands for finding which primaries and descendant fields are created by the short distance product of these two fields. These may be thought as a selection rules for the conformal dimensions of fields appearing in a thee point correlator. We say, for instance, that the fusion of $\Phi_{1}$ and $\Phi_{2}$ is possible if the three point function $\langle\Phi_{1} \Phi_{2}\Phi_{3}\rangle$ is not zero. Actually this problem is related to the presence of “null vectors” in the Virasoro representations, actually states $|\chi\rangle$
 obtained by action of $L_{n}$ with $n < 0$ on an highest weight, but such that $L_{m} |\chi\rangle = 0$ when $m > 0$ (we remark that such a pathology does not exist in usual simple algebra representations).

Considering the case of the critical Ising model, actually the simplest non trivial unitary minimal model, there are, in addition to the identity operator, two local 
scaling operators: the Ising spin $\sigma$ and the energy density $\epsilon$ (continuum version of lattice spin $\sigma_{i}$ and interaction energy $\sigma{i}\sigma_{i+1})$. These three operators can be identified with the minimal model $c=1/2$ and $h= 1/2, 1/16, 0$ with operator field correspondence: $\Phi$= $\epsilon$, $\sigma$, 1 satisfying the fusion rules: 
\[
\sigma\otimes\sigma= 1 + \epsilon \hspace{1,5cm}      \sigma\otimes\epsilon  = \sigma  \hspace{1,5cm}      \epsilon\otimes\epsilon=1 
\]
This algebra is isomorphic to the algebra ${ D, \eta, 1}$ of global symmetry of the critical Ising model: 
\[
D\otimes D = 1 +\eta   \hspace{1,5cm}    D\otimes \eta = D  \hspace{1,5cm}   \eta\otimes\eta = 1
\]
One can see that $\sigma$ (or $D$) is not invertible.

\section{An application of Quantum groups: a model for the Genetic Code}

There is a domain in the sciences of life particularly adapted to welcome methods and mathematical tools used in physics, and it is Genetics. Already about eighty years ago, Erwin Schrödinger provided in his book “\textsl{What is life?}” [90] some ideas about the possible role of a “\textsl{new physics}” in this domain, imagining for example mutations to be directly linked to quantum leads. As can be read there: 

\textsl{“ living matter, while not eluding the “laws of physics” as established up to date, is likely to involve “other laws of physics” hitherto unknown, which however, once they have been revealed, will form just as integral a part of science as the former.”}

Several attempts shown up in this direction these last years (as an example one can mention the p-adic analysis of the genetic code [91]). Symmetry appears also as a promising way for modelizing the genetic code, and it is this approach that we have started almost thirty years ago [92] and developed up to recently. As this part of the report deals with a different domain with respect to the subjects above considered, it may be useful to start with a short introduction on the genetic code. 
 
First, as well known, the DNA macromolecule is constituted by two chains of nucleotides wrapped in a double helix shape. There are four different nucleotides, characterized by their bases: adenine (A) and guanine (G) deriving from purine, and cytosine (C) and thymine (T) coming from pyrimidine. Note also the A (reps. T) base in one strand is connected with two hydrogen bonds to a T (resp. A) base in the other strand, while a C (resp. G) base is related to a G (reps. C) base with three hydrogen bonds: this is known as the complementary rule. The genetic information is transmitted to the cytoplasm via the messenger ribonucleic acid (mRNA). During this operation called transcription, the A, G, C, T bases in the DNA are associated respectively to the U, C, G, A bases, U denoting the uracile base. Then it will be through a ribosome that a triplet of nucleotides or codon will be related to an amino acid (a.a.). More precisely, a codon is defined as an ordered sequence of three nucleotides, e.g. AAG, ACG, etc., and one enumerates in this way 4 ×4×4 = 64 different codons. In the universal Eukariotic Code, 61 of such triplets can be connected in an unambiguous way to the amino-acids, except the three following triplets UAA, UAG and UGA, which are called non-sense or stop codons, the role of which is to stop the biosynthesis. Indeed the genetic code is the association between codons and amino-acids. But since one distinguishes only 20 amino-acids related to the 61 codons, it follows that the genetic code is degenerate. Still considering the standard Eukariotic Code, one observes sextets, quadruplets, triplet, doublets and singlet of codons, each multiplet corresponding to a specific amino-acid (a.a.).

In our model that we called the “Crystal Basis Model”, the four nucleotides as basic states of the (1/ 2, 1/ 2 ) representation of the $U_{q}(Sl(2) \oplus{Sl(2)})$ quantum enveloping algebra in the limit q=0 [90]. A triplet of nucleotides will then be obtained by constructing the tensor product of three such four-dimensional representations. Actually, this approach mimicks the group theoretical classification of baryons made out from three quarks in elementary particles physics, the building blocks being here the A, C, G, T/U nucleotides. The main and essential difference stands in the property of a codon to be an ordered set of three nucleotides: for example the codons (ACG) and (CGA) are two different codons. And that is not the case for a baryon, the wave function of which being a linear combination of states, each made of the three quarks constituting the hadron in a different order: as an example the proton made of two u quark and one d quark appears as a linear combination of  the states uud, udu and duu. Constructing such pure states is made possible in the framework of any algebra $U_{q=0}(G)$with $G$ being any (semi)-simple classical Lie algebra owing to the existence of a special basis, called crystal basis, in any (finite dimensional) representation of $G$. The algebra $G = Sl(2)\oplus{Sl(2)}$ appears the most natural for our purpose. The complementary rule in the DNA–mRNA transcription may suggest to assign a quantum number with opposite values to the couples (A,T/U) and (C,G). The distinction between the purine bases (A,G) and the pyrimidine ones (C,T/U) can be algebraically represented in an analogous way. Thus considering the fundamental representation (1/2 , 1/2 ) of $Sl(2)\oplus{Sl(2)}$ and denoting ± the basis vector corresponding to the eigenvalues ±1/2 of the $J_{3}$ generator in any of the two $Sl(2)$ corresponding algebras, we will assume the following “biological” spin structure: 

$\hspace{7cm}Sl(2)_{H}$

$\hspace{4cm}$ $C=(+,+)$ $\hspace{1cm}$ $\leftrightarrow$$\hspace{1cm}$ $U=(-,+)$

$\hspace{2.8cm}$ $Sl(2)_{V} $ $\updownarrow$ $\hspace{4.3cm}$    $\updownarrow$  $Sl(2)_{V}$

$\hspace{4cm}$ $G=(+,-)$ $\hspace{1cm}$ $\leftrightarrow$$\hspace{1cm}$ $A=(-,-)$

$\hspace{7cm}Sl(2)_{H}$

the subscripts $\textsl{H}$  (:= horizontal) and $\textsl{V}$ (:= vertical) being just added to specify the algebra. The method for performing the tensorial product of $U_{q=0}(G)$ representations has been established by Kashiwara [93] . 

Considering the three-fold tensor product, we will get, as for usual product of $Sl(2)\oplus{Sl(2})$ representations, the irreducible representations :
\[
 (1/2, 1/ 2) \otimes {(1/2 , 1/ 2 )} \otimes {(1/2 , 1/ 2 )} = (3/2 , 3/2 ) \oplus{ 2(3/2 , 1/2 )} \oplus{ 2(1/2 , 3/2 )}\oplus{  4(1/2 , 1/2 )}  
\]
but the codon content in each of the obtained irreducible representations is the following, the upper labels denoting different irreducible representations, with $SL(2)_{H}$ acting horizontally and  $SL(2)_{V}$ vertically:
\[
                          (3/2, 3/2)  =  \left(\begin{array}{cccc}
                     CCC&UCC&UUC&UUU\\
                     GCC&ACC&AUC&AUU\\
                     GGC&AGC&AAC&AAU\\
                     GGG&AGG&AAG&AAA
                    \end{array}\right)
\]
\[
(3/2,1/2)^{1} = \left(\begin{array}{cccc} 
                      CCG&UCG&UUG&UUA\\
                      GCG&ACG&AUG&AUA
                      \end{array}\right)
(3/2,1/2)^{2} =  \left(\begin{array}{cccc}  
                      CGC&UGC&UUG&UUA\\
                      CGG&UGG&UAG&UAA
                      \end{array}\right)
  \]                 
\[
(1/2,3/2)^{1} =  \left(\begin{array}{cc} 
                      CCU&UCU\\
                      GCU&ACU\\
                      GGU&AGU\\
                      GGA&AGA
                      \end{array}\right)
\qquad\qquad
(1/2,3/2)^{2} =  \left(\begin{array}{cc} 
                      CUC&CUU\\
                      GUC&GUU\\
                      GAC&GAU\\
                      GAG&GAA
                      \end{array}\right)
\]
\[
(1/2, 1/2)^{1} =   \left(\begin{array}{cc} 
                      CCA&UCA\\
                      GCA&ACA
                     \end{array}\right)
\qquad\qquad
(1/2, 1/2)^{2} =   \left(\begin{array}{cc} 
                      CGU&UGU\\
                      CGA&UGA
                     \end{array}\right)  
\]                
\[
(1/2, 1/2)^{3} =   \left(\begin{array}{cc} 
                      CUG&CUA\\
                      GUG&GUA
                     \end{array}\right)
\qquad\qquad
(1/2, 1/2)^{4} =   \left(\begin{array}{cc} 
                      CAC&CAU\\
                      CAG&CAA
                     \end{array}\right)
\]

With this “parameterisation” of the Genetic Code constituents, allowing us to propose computation of biological quantities and construction of operators adapted to represent biological situations. Hereafter is a rapid summary of the results we have obtained, the interested reader is invited to consult our review paper [94] (see also [95]).

As a first application is the setting of sum rules for codon usage probabilities : it is deduced that the sum of usage probabilities of codons with C and A in the third position for the quartets and/or sextets is independent of the biological species for vertebrates. More recently, using the codon-anticodon interaction potential mentioned below, an analysis of the codon usage frequencies of a specimen of 20 plants for which the codon-anticodon pattern is known, we have remarked that the hierarchy of the usage frequencies present an almost “universal” behaviour. 

A second application deals with the physical-chemical properties of amino-acids for which a set of relations have been derived and compared with the experimental data. A prediction for the not yet measured thermo-dynamical parameters of three amino-acids is also proposed. 

Another important notion in physics is the principle of minimal action, or use of minimum of energy. A codon-anticodon interaction potential is proposed in the framework of the model. Such an interaction can be compared with the spin-spin interaction on particle physics, the two $Sl(2)$ above defined being associated to two “biological” spin. Such a study will first allow to determine the structure of the minimum set of 22 anticodons allowing the translational-transcription for animal mitochondrial code. The results are in very good agreement with the observed anticodons. Then, the evolution of the genetic code is considered, with 20 amino-acids encoded from the beginning, from the viewpoint of codon-anticodon interaction. Following the same spirit as above, a determination of the structure of the anticodons in the Ancient, Archetypal and Early Genetic codes is obtained. Most of our results agree with the generally accepted scheme. Finally, keeping still at hand the minimization of our codon-anticodon interaction potential, codon bias are discussed, providing inequalities between codon usage probabilities for quartets of codons. Performing this study separately for the Early and for the Eukariotic genetic code, we observe a consistency with the obtained results as well as good agreement with the available data. Last but not least, an analysis of the coherent change of sign, in the evolution from the Early to the Eukaryotic code, of the two parameters regulating our interaction potential is performed. 

\section{Conclusion}

This report is a modest attempt for summarizing the development and the impact of symmetries in physics and even in genetics. As already mentioned in the course of these pages, it can be considered as a first step towards a more developed program on the “Symmetries in Modern Physics” [61]. \\
Let me conclude with the premonitory words of Gaston Bachelard: \textsl{“Soon, without doubt, abstract Physics will order all the possibilities of experiment”}. It seems to me that it is more or less the case these days, in particular with the help of symmetries…

\section{Acknowledgements}

I am grateful to Eric D’Hoker, Branko Dragovich, Pierre Fayet, Luc Frappat, Malte Henkel and Marios Petropoulos for helpful comments, informations and encouragements in the elaboration of this report. I also wish to thank the organizers of the perfect meeting they offered to all the participants. 

\section*{References}

[1] P. Ramond, “\emph{Introduction to Exceptional Lie groups and Algebras}” CALT-68-577

[2] J.M. Levy-Leblond, “\emph{Une nouvelle limite non-relativiste du groupe de Poincaré}”  Ann. Inst. H.~Poincaré, 2 (1965) 1.

[3] L. Ciambelli, R. G. Leigh, C. Marteau and P.M. Petropoulos, “\emph{Carroll Structures, Null Geometry and Conformal Theories}”, arXiv: hep-th/1905.02221; N. Athanasiou, P.M. Petropoulos, S. M. Schulz and G. Taujanskas “\emph{One dimensional Carrollian fluids. Part 1.Carroll-Galilei duality}” arXiv: hep/th 2407. 05962, ibid “\emph{One-dimensional Carrollian fluids II: C1 blow-up criteria}”. arXiv: hep/th 2407. 05971; P.M. Petropoulos, S. M. Schulz and G. Taujanskas “\emph{One-dimensional Carrollian fluids III: Global existence and weak continuity in $L^\infty$}” arXiv: hep/th 2407. 05972.

[4] L. Frappat, A. Sciarrino and P. Sorba, “\emph{Dictionary on Lie algebras and Superalgebras}”, Academic Press (2000)

[5] Juan M. Maldacena, “\emph{The large N limit of Superconformal Field Theories and Super gravity}”, arXiv: hep-th/9711200

[6] O. Aharoni, O. Bergman, D. L. Jafferis, J. Maldacena, “\emph{N=6 superconformal Chern-Simons-matter theories, M2-branes and their gravity duals}”, arXiv : hep/th: 0806.1218

[7] P.K. Kovtun, D.T. Son and A.O. Starinets, “\emph{Viscosity in strongly interacting quantum field theories from black hole physics}”,  arXiv: hep/th: 0405231 

[8] Sachdev, Subir (2013). "\emph{Strange and stringy}". Scientific American. 308 (44): 44–51.  

[9] H. Bondi, M.G.J. van der Burg and A.W. K. Metzner, “\emph{Gravitational waves in general relativity. 7. Waves from antisymmetric isolated systems}” Proc.Roy.Soc. Lond. A269 (1962) 21-52; R.K. Sachs, “\emph{Gravitational waves in general relativity.8.Waves in asymptotically flat space-times}”, Proc.Roy.Soc. Lond. A270 (1962) 103-126; id.\emph{“Asymptotic symmetries in gravitational theory}” Phys.rev.128 (1962) 2851-2864.

[10] R. Ruzziconi, “\emph{On the various extensions of the BMS group}” Thesis, Bruxelles (2020) hep-th: 2009.01926

[11] G. Barnich and C. Trossaert, “\emph{Symmetries of asymptotically flat 4-dimensional space-times at null infinities revisited}” Phys.Rev.Lett.105 (2010)111103, hep-th: 0909.2617; “\emph{Supertranslations call for superrotations}”hep-th: 1102-4632, and references therein.

[12] A.A. Belavin, A.M. Polyakov and A.B. Zamolodchikov, “I\emph{nfinite conformal symmetry in two-dimensional field theory}” Nucl.Phys. B241, 333 (1984).

[13] C. Duval, G.W. Gibbons and P.A. Horvathy, “\emph{Conformal Carroll groups and BMS symmetry}” hep-th: 1402.5894

[14] C. Duval, G.W. Gibbons and P.A. Horvathy, “\emph{Conformal Carroll groups}” hep-th: 1403.4213

[15] U. Niederer, “\emph{The Maximal Kinematical invariance Group of the Free Schrödinger Equation}” Helv.Phys.Acta, 45 (1972) 802

[16] G. Burdet, M. Perrin and P. Sorba, “\emph{About the Non-relativistic structure of the Conformal algebra}” Comm.Math.Phys. 34 (1973) 85  

[17] M. Henkel,   “\emph{Schrödinger invariance and strongly anisotropic critical systems}” J.Stat.Phys.75 (1994) 1023-1061

[18] M. Henkel, “\emph{Phenomenology of local scale invariance:from conformal invariance to dynamical scaling”}, Nucl.Phys. B 641 (2002) 405-486; “\emph{Ageing, dynamical scaling and conformal invariance}”, Int.J.Mod.Phys. A 19 (2004) 2207-2216.

[19] C. Duval, M. Henkel, P.A. Horvathy, S. Rouhani and P.M. Zhang, “\emph{Schrödinger Symmetry : an historical review”} , arXiv: 2403.20316v3 hep-th/ 19 May 2025

[20] Juan Maldacena, Dario Martelli and Yuji Tachikawa, “\emph{Comments on string theory backgrounds with non relativistic conformal symmetry}”, arXiv: hep-th/0807.1100

[21] Sean Hartnoll, \emph{“Lectures on Holographic methods for Condensed Matter Physic}s”, arXiv: hep-th/0903.3246 V3

[22] Y. Nishida and D.T. Son, “\emph{Non relativistic conformal field theories} “ arXiv : 0706.3746 ; D.T. Son “ \emph{Toward an AdS/cold atoms correspondence”} arXiv: hep-th/0804.3972

[23] A. Bagchi and R. Gopakumar, \emph{“Galilean conformal algebras and AdS/CFT}”arXiv: hep/th 0902.1385V3

[24] C. Duval and S. Lazzarini, “\emph{Schrödinger manifolds”} math-ph: 1201.0683

[25] E.A. Bergshoeff, J.M. Figueroa-O’Farrill and J. Gomis, “\emph{A Non-Lorentzian Primer}”hep-th: 2206.12177

[26] see M. Gell-Mann and Y. Ne’eman “\emph{The Eightfold Way}”, W.A. Benjamin, New-York, 1964.

[27] M. Gell-Mann, “\emph{A schematic model of Baryons and Mesons}”, Phys.Lett.8 (3) 214-215 ; G. Zweig, “\emph{An SU(3) model for Strong Interaction Symmetry and its Breaking}” CERN-TH-401 and CERN-TH-412.

[28] M.Y. Han and Y. Nambu, \emph{“Three-Triplet with double SU(3) Symmetry}”Phys.rev.B139 (4B) :1006 (1965)

[29] H. Fritzsch, M. Gell-Mann and H. Leutwyler, “\emph{Advantages of the Color octet gluon picture}”, Phys.Lett.B47 (4), 365-368 (1973) 

[30] H. Hogaasen and P. Sorba, “\emph{On Quarks and Flavour Symmetry”}, hep-th: 0711.3157  

[31] R.J. Jaffe, “\emph{Multiquark Hadrons 1, The Phenomenology of q2q*2 Mesons”} Phys.Lett.D15 (1977) 267; id. “\emph{Multiquark Hadrons 2. Methods}”, Phys.Rev.D15 (1977) 281.

[32] V.A. Matveev and P. Sorba, “\emph{Is deuteron a six quark system?}” Lett.Nuovo Cim.20 (1977) 435

[33] V.A. Matveev and P. Sorba “\emph{Quark analysis of multibaryonic systems}” Nuovo Cim.A45, 257 (1978); H. Hogaasen, P. Sorba and R. Viollier, Z. Physik C4, 131 (1980)

[34] S. Kaur, J. Wu, S. Xu, C. Mondal, X. Zhao, J.P. Vary, \emph{“Color structure of deuteron on the light front}” hep-th :2503.21371;  S. Kaur, C. Mondal, X. Zhao, C.-R. Ji, “\emph{Structure of lightest nuclei in the visible Universe}” hep-th :2507.09886;  

[35] H.-M. Chan and H. Hogaasen, \emph{“Baryonium states in multiquark spectroscopy”},Phys. Lett.B72 (1977) 121; id. “\emph{A model for baryonium”}, Phys.Lett. B136 (1978) 634.

[36] H. Hogaasen and P. Sorba, “\emph{The systematic of possibly narrow states with baryon number one”} Nucl.Phys.B145 (1978) 119; M. De Combrugghe, H. Hogaasen and P. Sorba, “\emph{Narrow multiquark baryons}”, Nuclear Phys.B156 (1979), 261.

[37] H. Hogaasen and P. Sorba, “\emph{What is happening to multiquark baryons}?”
Proc.of the 4th Intern.Conf. on Baryon resonances “ (1980),p.401, Toronto, Canada.

[38] A. de Rujula, H. Georgi and S.L. Glashow, “\emph{Hadron masses in a gauge theory}” Phys.Rev. D12 (1975) 147

[39] C. Gignoux, B. Sylvestre-Brac and J.M. Richard, “\emph{Possibility of stable multiquark baryons}”, Phys.Lett.B193 (1987) 323.

[40] H. Hogaasen and P. Sorba, “\emph{On the discrete charm of multiquark states}” 
July 2020, unpublished.

[41] H. Hogaasen, J.M. Richard and P. Sorba, “\emph{A chromomagnetic mechanism for the X(3872) resonance}”,hep-th/0511039; [B\&H\&R\&S]

[42] F. Buccella, H. Hogaasen, J.M. Richard and P. Sorba, “ \emph{Chromomagnetism, flavour symmetry breaking and S-wave tetraquarks}”, hep-th/ 0608001

[43] LHCb Collab., R. Aaij et al , hep-ex/1507.03414                  

[44] LHCb Collab., R. Aaij et al , hep-ex/ 2006.16957

[45] F. Buccella \emph{“b decays: a factory for hidden charm multiquark states}” hep-th/ 1801.03723

[46] C.N. Yang and R. Mills, “\emph{Conservation of Isotopic Spin and Isotopic Gauge Invariance}”, Phys.Rev.96 (1954) 19

[47] S. Glashow "\emph{The renormalizability of vector meson interactions}", Nucl. Phys. 10, 107 (1959);  A.Salam, in Elementary Particle Theory, ed.N.Swartholm (Almquist and Forlag,Stockholm, 1968; S. Weinberg, \emph{“A model of Leptons}”, Phys. Rev. Letters 19 (21): 1264–66

[48] P.W. Higgs, “\emph{Broken symmetries and the mass of the Gauge Bosons}” Phys.Rev.Lett. 12 (1964) 132; F. Englert and R. Brout, “B\emph{roken Symmetry and the Mass of Gauge Vector Mesons}”, Phys.Rev.Letters, 13 (1964) 321; 

[49] A. Neveu and Schwarz, Nucl.Phys.B31(1971) 86; P. Ramond, Phys.Rev.D3 (1971) 2415; Y. Aharonov, A. Casher and L. Susskind, Phys.Lett.B35 (1971) 512; J.L. Gervais and B. Sakita, Nucl.Phys.B34 (1971) 633.

[50] J. Wess and B. Zumino, \emph{“Supergauge transformations in four dimensions}”Nucl.phys.B70 (1974) 39

[51] S. Coleman and J. Mandula, “\emph{All possible symmetries of the S-matrix”,} Phys.rev. 159 (1967) 1251

[52] Yu.A. Golfand and E.P. Likhtmann, “\emph{Extension of the algebra of the Poincaré group generators and violation of P invariance”}, JETP Lett.13 (1971) 323.

[53] R. Haag, J.T. Lopuszanski, M.F. Sohnius “\emph{All possible generators of supersymmetries of the S-Matrix}”, Nucl.Phys.B88 (1975) 257

[54] P. Fayet and S. Ferrara, “\emph{Supersymmetry}”,Phys.Reports 32C (1977) 249

[55] M.F. Sohnius, “\emph{Introducing Supersymmetry}”, Phys.Reports 128 (1985) 39

[56] A.S. Galperin, E.A. Ivanov, V.I. Ogievetsky and E.S. Sokatchev, \emph{“Harmonic superspace”} (2001) Cambridge University Press, p.306  

[57] P. Fayet, “\emph{Supergauge invariant extension of the Higgs mechanism and a model for the electron and its neutrino”} Nucl.Phys. B90 (1975) 104-124

[58] P. Fayet, “\emph{Fermi-Bose-Hypersymmetry” }Nucl.Phys.B113(1976) 135 P. Fayet, “\emph{Spontaneous generation of massive multiplets and central charges in extended supersymmetric theories}” Nucl. Phys. B149 (1979) 137-169

[59] J. Wess and J. Bagger, “\emph{Supersymmetry and Supergravity}”, Princeton University Press, 1992 

[60] V. Kac, “\emph{Classification of Lie Superalgebras”}, Funct. Anal. Appl. 9(1975)263; “\emph{Lie Superalgebras}”, Adv.Math.Phys. 26(1977)8; “\emph{A sketch on Lie Superalgebra theory”}, Comm. Math.Phys.53 (1977) 31; “\emph{Characters of typical representations of classical Lie Superalgebras}”, Comm. Math.Phys.5(1977) 889; “\emph{Representations of classical Lie Superalgebras”,} Lecture Notes in Mathematics 676 (1978) 597 Springer-Verlag, Berlin.

[61] E. D’Hoker and P. Sorba, “\emph{Symmetries in Modern Physics}” in preparation.

[62] P.G.O. Freund and M.A. Rubin, “\emph{Dynamics of Dimensional Reduction}” Phys.Lett. B97, (1980) 233; L.J. Romans, “\emph{New compactification of chiral N=2 D=10 Supergravity}”, Phys.Lett.B153 (1985) 392; J. Figueroa-O’Farrill and G. Papadopoulos, “\emph{Maximally supersymmetric solutions of ten and eleven dimensional supergravities}”, hep-th/0211089.

[63] K. Becker, M. Becker and J. Schwarz, “\emph{String theory and M-theory: a modern introduction}”, Cambridge (2007).

[64] E. D’Hoker and D.Z. Freedman, \emph{“Supersymmetric Gauge Theories and the ADS/CFT Correspondence}”, hep-th / 0201253

[65] D. Andriot, \emph{“Dark energy from String Theory: an introductory review”} hep-th/2603.2579

[66] M.B. Green and J. Schwarz, “\emph{Anomaly cancellations in supersymmetric D=10 gauge theory and superstring theory}” Phys. Lett. 149 (1984) 11

[67] Ed. Witten, “\emph{String theories dynamics in various dimensions} “ hep-th/ 9503124

[68] Ph. Di Francesco, P. Mathieu and D. Senechal, “\emph{Conformal Field Theory}”, Springer (1997)

[69] A. Cappelli, C. Itzykson and J.-B. Zuber, “\emph{Modular invariant partition functions in two dimensions”}, Nucl.Phys.B280 (1987) 445; “\emph{The A-D-E classification of minimal and A1(1)conformal invariant theories”,} Comm.Math.Phys.13 (1987) 1

[70] P. Goddard and D. Olive, “\emph{Kac-Moody and Virasoro algebras in relation to Quantum Physics}”, Int.Journ.of Mod.Phys.A, Vol.1, No 2 (1986) 303-414
\emph{
[71] D. Lust and S. Theisen, “}Lectures on String Theory”, Lecture Notes in Physics 346 (1989) Springer-Verlag

[72] Ed. Witten,”\emph{Non Abelian Bosonisation in two dimensions}” Comm.Math.Phys.92 (1984) 455; V.G. Knizhnik and A.B. Zamolodchikov, “\emph{Current Algebra and Wess-Zumino models in two dimensions}”Nucl.Phys. B247 (1984) 83; D. Gepner and E. Witten, “\emph{String Theory on Group Manifolds}” Nucl.Phys.B278 (1986) 493” 

[73] L. Frappat, A. Sciarrino and P. Sorba, \emph{“Structure of basic superalgebras and of their affine extensio}ns”, Comm.Math.Phys.121 (1989) 457

[74] E. Ragoucy and P. Sorba, \emph{“Extended Kac-Moody algebras and applications”}, Intern.Journ.of Mod.Phys.A ,Vol7, No13 (1992) 2883-2971

[75] R. Campoamor-Stursberg, A. Marrani  and M. Rausch de Traubenberg, “\emph{New perspectives in Kac-Moody algebras associated to higher dimensional manifolds}” , hep-th/2510.02953  and ref.therein

[76] A.N. Leznov and M.V. Saveliev, \emph{“Exactly and Completely Integrable Nonlinear Dynamical Systems”}, Acta Appl.Math.116 (1989) and ref. therein

[77] L. Feher, L. O’Raifeartaigh, P. Ruelle, I. Tsutsui and A. Wipf,\emph{”On Hamiltonian reduction of the Wess-Zumino-Novikov-Witten theories”} Phys.Rep. 222 (1992)1, and ref.therein

[78] F. Delduc, L. Frappat, E. Ragoucy and P. Sorba, \emph{“General properties of Classical W algebras”}, hep-th/931204

[79] A.N. Leznov and M.V. Saveliev, \emph{“Representations of zero curvature for the system of nonlinear partial differential equations $x_{\alpha,zz} = exp(kx)_{\alpha}$ and its integrability”}, Lett.Math.Phys.3 (1979) 485

[80] E. D’Hoker and D.H. Phong, “\emph{Lectures on Supersymmetric Yang-Mills Theory and Integrable Systems}”, hep-th/9912271

[81] N. Seiberg and E. Witten, “\emph{Electro-magnetic duality, monopole condensation, and confinement in N = 2 supersymmetric Yang-Mills theory}”, hep-th/9407087; “\emph{Monopoles, duality, and chiral symmetry breaking in N = 2 supersymmetric QCD”}, hep-th/9410167.

[82] F. Calogero, “\emph{Exactly solvable one-dimensional many-body problems}”, Lett. Nuovo Cim. 13 (1975) 411-416; J. Moser, “\emph{Three integrable Hamiltonian systems connected with isospectral deformations}”, Advances Math. 16 (1975) 197

[83] V. Chari and A. Pressley, “\emph{A guide to Quantum Groups}”, Cambridge University Press (1994)

[84] R.J. Baxter, “\emph{Exactly Solved Models in Statistical Mechanics}”. Academic Press” (1982)

[85] P. Sorba, \emph{“Deformed Virasoro and W algebras from elliptic algebras”}, Max-Planck Institute für Mathematik Preprint MPI 99-39 (1999).

[86] D. Gaiotto, A. Kapustin, N. Seiberg and B. Willett, “\emph{Generalized Global Symmetries”} hep-th/ 1412.5148

[87] E. Lake,“\emph{Higher-form Symmetries and Spontaneous Symmetry Breaking}”, hep-th/1802.07747 ;  P. Gomez, “ \emph{An Introduction to Higher Form Symmetries}”, hep-th/2303. 01817; L. Bhardwaj et al, “\emph{Lectures on Generalized Symmetries”, }hep-th/ 2307.07547; J. McGreevy, “\emph{Generalized Symmmetries in Condensed Matter}”, hep-th/ 2204.03045 

[88] C. Delcamp, E. Heng and M. Yu, \emph{“A non-semisimple non-invertible symmetry}” hep-th/2412.19635 

[89] C. Chang, Y. Lin, S. Shao, Y. Wang and X. Yin, “\emph{Topological Defect Lines and renormalization Group Flows in Two Dimensions}” , hep-th/ 1802.04445

[90] E. Schrödinger, \emph{“What is life?”}, Cambridge University Press (1944)

[91] B. Dragovich and A. Dragovich,  “\emph{p-Adic Modelling of the Genome and the Genetic Code}” arXiv:0707.3043 [q-bio.OT].

[92] F. Frappat, A. Sciarrino and P. Sorba, “\emph{A Crystal Basis for the Genetic Code”,} Phys. Lett. A 250 (1998) 214; “Symmetry and Codon Usage correlations in the Genetic Code”, Phys. Lett. A 259 (1999)339.

[93] M. Kashiwara, Comm.Math.Phys. 133, 249 (1990)

[94] A. Sciarrino and P. Sorba, “\emph{Symmetry and Minimum Principle at the basis of the Genetic Code}”, arXiv: 1704.0094 [q-bio.OT]

[95] A. Sciarrino and P. Sorba, “\emph{Hierarchy of codon usage frequencies from codon-anticodon interaction in the crystal basis model}”, arXiv: 2312.11107 [q-bio.OT]

\end{document}